\documentclass{article}
\usepackage{spconf,amsmath,graphicx}
\usepackage{amssymb}
\usepackage{multirow}
\usepackage[colorlinks,linkcolor=blue]{hyperref}
\usepackage{wrapfig}

\title{AAT: Adapting Audio Transformer for Various\\Acoustics Recognition Tasks}
\name{Yun Liang$^{1, 2}$, Hai Lin$^{1}$, Shaojian Qiu$^{1, 2, \star}$, Yihang Zhang$^{1}$\thanks{$^{\star}$Corresponding Author: Shaojian Qiu 
\\First Author and Second Author contribute equally to this work.\\
This study was funded partly by key R\&D project of Guangzhou (202206010091, 2023B03J1363), the Science and Technology Planning Project of Guangdong Province (2019A050510034), Meizhou Tobacco Technology Project of Guangdong Province (202304), Guangdong Natural Science Fund Project (2022A1515110564), Guangzhou Science and Technology Plan Project (202201010312).
}}
\address{$^{1}$South China Agricultural University, Guangzhou, China\\$^{2}$Guangzhou Key Laboratory of Intelligent Agriculture, Guangzhou, China}
\begin{document}
\ninept
\maketitle
\begin{abstract}
Recently, Transformers have been introduced into the field of acoustics recognition. They are pre-trained on large-scale datasets using methods such as supervised learning and semi-supervised learning, demonstrating robust generality——It fine-tunes easily to downstream tasks and shows more robust performance. However, the predominant fine-tuning method currently used is still full fine-tuning, which involves updating all parameters during training. This not only incurs significant memory usage and time costs but also compromises the model's generality. Other fine-tuning methods either struggle to address this issue or fail to achieve matching performance. Therefore, we conducted a comprehensive analysis of existing fine-tuning methods and proposed an efficient fine-tuning approach based on Adapter tuning, namely AAT. The core idea is to freeze the audio Transformer model and insert extra learnable Adapters, efficiently acquiring downstream task knowledge without compromising the model's original generality. Extensive experiments have shown that our method achieves performance comparable to or even superior to full fine-tuning while optimizing only 7.118\% of the parameters. It also demonstrates superiority over other fine-tuning methods. 
% Code is available at \url{https://github.com/MichaelLynn1996/AAT}.
\end{abstract}
\begin{keywords}
Acoustics recognition, pre-trained model, Adapter, Prompt, parameter-efficiency fine-tuning
\end{keywords}
\section{Introduction}
\label{sec:intro}
% 背景
In recent times, Transformer-based deep neural networks, which rely on multi-head self-attention (MHSA) mechanisms, have been introduced into the field of acoustics recognition \cite{gong21b_interspeech, chen2022hts, koutini2021efficient}. They have been trained on large-scale acoustics datasets \cite{gemmeke2017audio} through supervised learning, exceeding CNN-based methods \cite{kong2020panns, gong2021psla} and consistently producing reliable results. Additionally, many efforts have been made to construct self-supervised learning frameworks to train audio Transformers on extensive unlabeled data, further uncovering the model's representation learning capabilities \cite{gong2022ssast, baade22_interspeech, huang2022masked}. However, the effective transfer of pre-trained audio Transformers from large-scale datasets to a variety of downstream tasks remains an unresolved issue.

\begin{figure}[t]
  \centering
  \includegraphics[width=\linewidth]{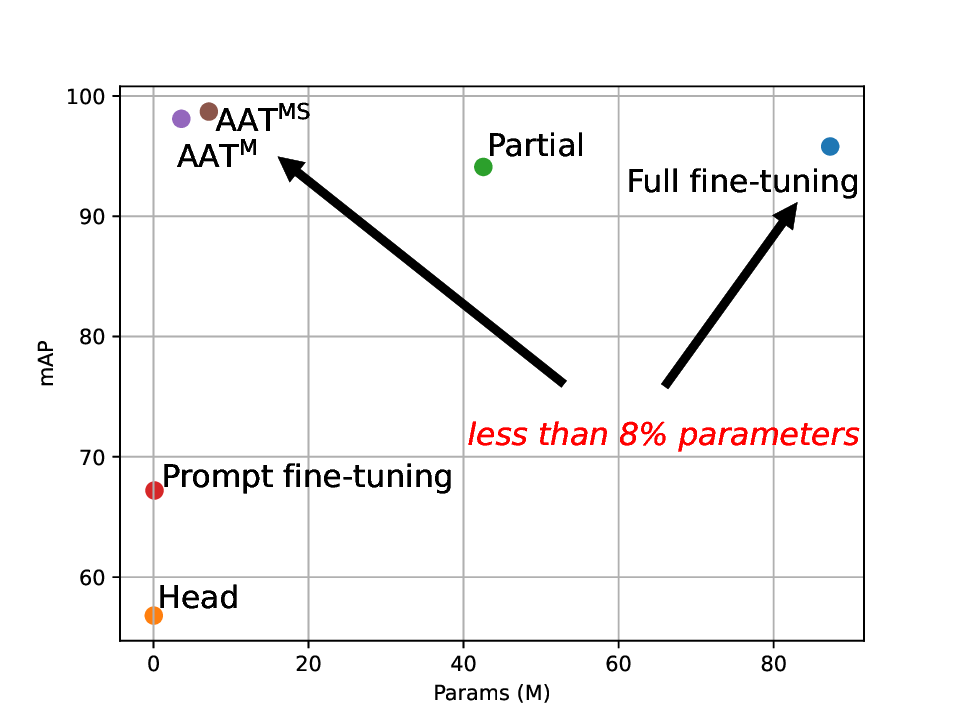}
  \caption{Performance comparison on Openmic dataset. Our proposed AAT achieves the highest accuracy while enjoying a significantly smaller number of tuning parameters.}
  \label{fig: trade-off}
\end{figure}
% 近况
Currently, the most direct approach is full fine-tuning \cite{gong21b_interspeech, chen2022hts, koutini2021efficient, gong2022ssast, baade22_interspeech, huang2022masked}, where all model parameters are updated during the back-propagation process. However, full fine-tuning comes with significant memory consumption and training time costs. Additionally, full fine-tuning adjusts the model's parameters entirely to a specific task, potentially leading to a loss of model generalization. This means that the fine-tuned model may perform poorly on other tasks because it has been optimized for a particular task. Freezing some network parameters while training others is another common fine-tuning method \cite{elshaer2019transfer, yosinski2014transferable}. This reduces the training cost substantially while achieving decent performance. However, partial fine-tuning still modifies some pre-trained model parameters, potentially compromising its generalization. Fine-tuning only the task-specific head \cite{kong2020panns} retains the model's original generalization. However, its performance can be significantly limited when the target task has a substantially different data distribution from the pre-training task.

% 问题
To address the limitations of the fine-tuning methods mentioned above, the parameter-efficient fine-tuning (PEFT) technique, has been extensively studied in the fields of natural language processing (NLP) \cite{houlsby2019parameter, liu2023pre, lester2021power} and computer vision (CV) \cite{chen2022adaptformer, jia2022vpt, yang2022aim}. However, these methods have seen limited exploration in the field of audio recognition. The core idea behind PEFT is to freeze the parameters of a pre-trained model and introduce additional parameters for fine-tuning. Two commonly used PEFT techniques are Adapter and Prompt tuning. Adapters \cite{chen2022adaptformer, houlsby2019parameter, yang2022aim} involve inserting a bottleneck module into the Transformer Encoder, adapting from the structural aspects of the model. Prompt tuning \cite{jia2022vpt, liu2023pre, lester2021power} entails feeding additional trainable tokens along with input embeddings into the Transformer encoders, it involves transferring from the input dimensionality of the model. These methods preserve the generality of the pre-trained model, save computational resources, and reduce data requirements. As a result, this has inspired us to utilize PEFT techniques to effectively transfer pre-trained audio Transformers to various downstream tasks in audio recognition.

% 提出方法
In this work, we propose a method to efficiently \textbf{A}dapt pre-trained \textbf{A}udio \textbf{T}ransformer models (AAT) for various acoustics recognition tasks. By freezing the pre-trained audio Transformer and adding a few lightweight Adapters during fine-tuning, we show that our proposed AAT can achieve competitive or even better results than full fine-tuning with substantially fewer tuning parameters, as shown in Fig. \ref{fig: trade-off}. To be specific, we first add a trainable Adapter without a shortcut namely \textbf{MLP Adapter} in parallel to the MLP layer in a Transformer block. The frozen MLP layer generates generic features, the MLP Adapter produces task-specific features, and the parallel design leads to a better fusion of these features. After that, we introduce a \textbf{Spatial Adapter} after the MHSA layer in a Transformer block to perform adaptation to feature space variations brought about by different sample lengths in various acoustics tasks.

% The key contributions of this paper are summarized as follows: (1) We propose a simple yet effective framework, namely AAT, for adapting audio Transformers to a large variety of downstream acoustics recognition tasks and keeping the generalization of the pre-trained model. (2) We compare various fine-tuning methods and demonstrate the superior robustness of AAT. (3) Extensive experiments on various downstream tasks demonstrate the effectiveness of AAT on multiple acoustics benchmarks, and our method is significantly more efficient than full fine-tuning.

\begin{figure}[h]
  \centering
  \includegraphics[width=\linewidth]{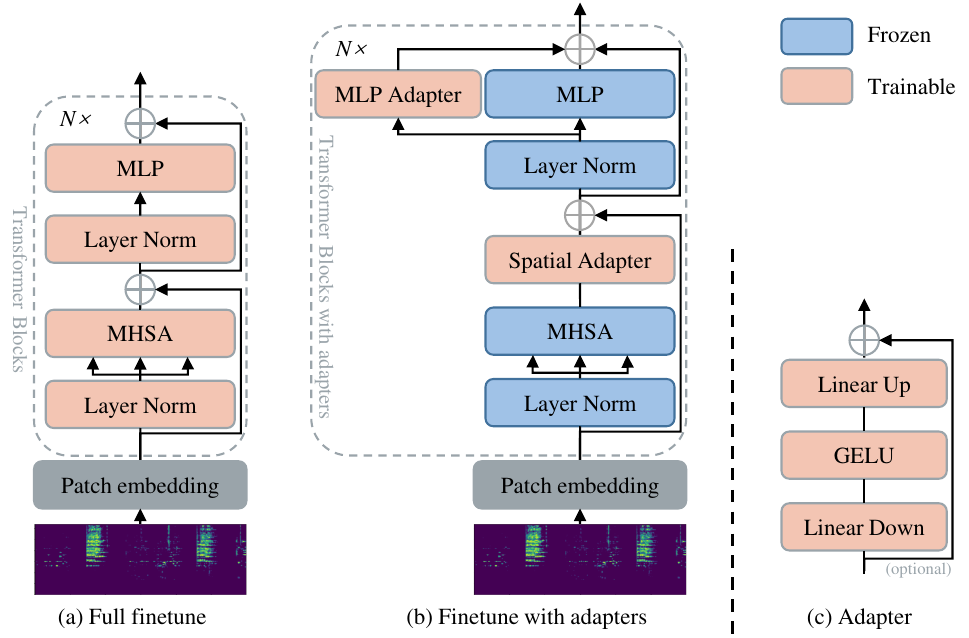}
  \caption{Brief illustration of full fine-tuning (a) and fine-tuning with Adapters (b) on pre-trained audio Transformer. (c) The architecture of the proposed Adapter.}
   \label{Overview}
\end{figure}

\section{Proposed method}
\label{sec:method}

We propose AAT for efficiently transferring large pre-trained audio Transformer models to downstream tasks. AAT attains strong transfer learning abilities by only fine-tuning a small number of extra parameters, circumventing catastrophic interference among tasks. We illustrate the overall framework of AAT in Fig. \ref{Overview}.

\subsection{Preliminary: Audio Transformer Architecture}

Transformer architecture was first introduced by \cite{gong21b_interspeech} into acoustics recognition. A vanilla audio Transformer basically consists of a patch embedding layer and several consecutively connected Transformer blocks, as depicted in Fig. \ref{Overview} (a). It takes spectrogram $x \in \mathbb{R}^{T \times F}$ as input, the patch embedding layer first splits and flattens the sample $x$ into sequential patches $x_p \in \mathbb{R}^{N \times (P^2d)}$, where $(T, F)$ represents the temporal dimension and frequency dimension of the input spectrogram, $(P, P)$ is the resolution of each spectrogram patch, $d$ denotes the output channel, and $N = HW /P ^2$ is the number of spectrogram tokens. The overall combination of a prepended $\mathrm{[CLS]}$ token and the spectrogram tokens $x_p$ are further fed into Transformer encoders for attention calculation.

Each Transformer block is composed of an MHSA and an MLP layer, together with Layer Norm (LN) \cite{ba2016layer} and skip connections, see Fig. \ref{Overview} (a). The computation of a MHSA layer can be written as:
\begin{equation}
{x'}_i=x_{i-1}+\operatorname{MHSA}(\operatorname{LN}(x_{i-1}))
\end{equation}

% \begin{equation}
% {x'}_i=\operatorname{Attention}(Q, K, V)=\operatorname{Softmax}\left(\frac{Q_i K^{T}}{\sqrt{d}}\right) V
% \end{equation}

\noindent where ${x'}_i$ are the tokens produced by MHSA at the $i$-th layer. The tokens ${x'}_i$ are further sent to a LN and a MLP block, which consists of two fully connected layers with a GELU activation function \cite{hendrycks2016gaussian} in between. This process is formally formulated as follows,

\begin{equation}
x_i={x'}_i+\operatorname{MLP}(\operatorname{LN}({x'}_i))
\end{equation}
where $x_i$ is the output of the $i$-th Transformer block. After the last Transformer block, the $\mathrm{[CLS]}$ token is sent to the task-specific head for the final classification. 
% In our work, firstly, we introduce an MLP Adapter in parallel to the MLP layer, and then we further add another Adapter with the skip connection for spatial Adaptation.
% replace the MLP layer with our AdaptMLP module for efficient fine-tuning purposes.

\subsection{AAT}

Inspired by PEFT techniques in NLP and CV, we designed our Adapter structure which can be shown in Fig. \ref{Overview} (c). It is simple yet efficient. The proposed Adapter is a bottleneck architecture that consists of two linear layers and a GELU in the middle. The linear down layer $W_\downarrow \in \mathbb{R}^{d \times \hat{d}}$ projects the input to a lower dimension and the linear up layer $W_\uparrow \in \mathbb{R}^{\hat{d}\times d }$ projects it to the original dimension.

% \begin{equation}
% \operatorname{Adapter}(z)=\operatorname{GELU}(z \cdot W_\downarrow) W_\uparrow
% \end{equation}
% To adapt the pre-trained spatial features to target video data, we add an Adapter after the self-attention layer as shown in Fig. 2(c), which we term as spatial adaptation. During training, all the other layers of the transformer model are frozen while only the Adapters are updated.

\subsubsection{MLP Adapter}

From the earlier review of audio Transformers, it can be summarized that an MLP layer often follows each MHSA layer. This is because the MLP layer prevents Transformers from degradation by preventing the MHSA from producing rank-1 matrices. Therefore, the MLP layer is necessary and crucial for Transformers \cite{dong2021attention}. Simultaneously, prior work \cite{szegedy2015going} has demonstrated that a parallel design is a more effective way of feature fusion. Hence, we initially introduced an Adapter without a shortcut in parallel with the MLP layer. During fine-tuning, the frozen MLP layer generates generic features, while the trainable Adapter produces task-specific features, leading to a better fusion of these features. We denote the variant of fine-tuning where only the MLP is fine-tuned as $\mathrm{AAT^{M}}$.

\subsubsection{Spatial Adapter}

Acoustics tasks often involve variable-length data. Audio Transformers are typically pre-trained on large-scale datasets with samples of 10 seconds in length. However, downstream task data can vary from 1 second (e.g. Speech Command) to 30 seconds (e.g. GTZAN), resulting in significant differences in spatial information between the downstream task data and the pre-trained audio Transformer data. This can impact the pre-trained audio Transformer's ability to capture global spatial information effectively. To address this issue, we introduce an additional Adapter with a shortcut after the MHSA to adapt to these spatial information variations. Due to the presence of shortcuts and zero initialization, the Spatial Adapter gradually becomes effective after a certain period of training. We denote the variant of fine-tuning where both the MLP and spatial domain are fine-tuned simultaneously as $\mathrm{AAT^{MS}}$.

The final structure of a Transformer block in our proposed AAT is presented in Fig. \ref{Overview} (b). The adapted procession can be written as:
\begin{equation}
{x}^S_i=x_{i-1}+\operatorname{Spatial\_Adapter}(\operatorname{MHSA}(\operatorname{LN}(x_{i-1})))
\end{equation}
\begin{equation}
x_i={x}_i^S+\operatorname{MLP\_Adapter}(\operatorname{LN}({x}^S_i))+\operatorname{MLP}(\operatorname{LN}({x}^S_i))
\end{equation}
where ${x}^S_i$ is the output of the MHSA layer with a Spatial Adapter and $x_i$ is the output of the MLP layer with a MLP Adapter.

\section{Experiments and results}
\label{sec:experiments}

\subsection{Experimental Settings}

\subsubsection{Pre-trained backbone}

% We adopt the Audio Spectrogram Transformer (AST) \cite{gong21b_interspeech} and Self-Supervised Audio Spectrogram Transformer (SSAST) \cite{gong2022ssast} as our backbone models, which pre-train by supervised learning (SL) and self-supervised learning (SSL) training methods, respectively. 
% Meaning they use the same model architecture but different pretraining methods.
We adopt the Audio Spectrogram Transformer (AST) \cite{gong21b_interspeech} which pre-train by supervised learning (SL) and Self-Supervised Audio Spectrogram Transformer (SSAST) \cite{gong2022ssast} which pre-train by self-supervised learning (SSL) as our backbone models.

\subsubsection{Initialization of weight}

For the original networks, we directly load the weights pre-trained on the upstream tasks and keep them frozen/untouched during the fine-tuning process. For the newly added Adapters, the weights of the linear down layer are randomly initialized, while the biases of the additional networks and the weights of the linear up layer are configured with zero initialization. In this way, the adapted model is close to the pre-trained model at the beginning of training, and the adapters gradually come into play during the parameter updates.

\subsubsection{Baseline methods}
\begin{wrapfigure}{r}{0.36\linewidth}
  \begin{center}
    \includegraphics[width=\linewidth]{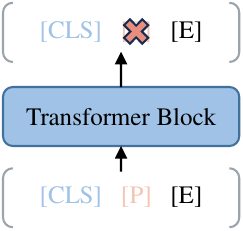}
  \end{center}
  \caption{The procession of Prompt tuning. $\mathrm{[CLS]}$ represent class tokens. $\mathrm{[P]}$ represent trainable Prompt tokens. $\mathrm{[E]}$ represent input spectrogram embeddings.}
  \label{Prompt}
\end{wrapfigure}
We selected four common fine-tuning methods as baselines for comparison with AAT, including:\\(1) Full: fully update all parameters of the pre-trained model.\\(2) Head: only update the task-specific head, which is a combination of a Layer Norm and a linear layer.\\(3) Partial: fine-tune the last half of the parameters of Transformer blocks within the backbone while freezing the others, as adopted in \cite{yosinski2014transferable}.\\(4) Prompt: fine-tune the extra prompt tokens parameters as shown in Fig. \ref{Prompt}. Prompt tokens $\mathrm{[P]}$ are added to the input spectrogram embedding $\mathrm{[E]}$ before being inputted into each Transformer block, and are removed when it produces its output, as adopted in \cite{jia2022vpt}. The number of Prompt tokens for each layer is set to 12.
% \begin{figure}[h]
%   \centering
%   \includegraphics[width=0.4\linewidth]{figures/Prompt.pdf}
%   \caption{The procession of Prompt tuning. $\mathrm{[CLS]}$ is class tokens. $\mathrm{[P]}$ is Prompt tokens. $\mathrm{[E]}$ is spectrogram embeddings.}
%    \label{Prompt}
% \end{figure}

% \vspace*{-10pt}
\subsubsection{Downstream tasks and implementation details}

We conducted experiments on six datasets representing three major categories: (1) \emph{Event} classification: ESC-50 \cite{piczak2015dataset} (ESC) for environmental sound classification and UrbanSound8k \cite{urbansound} (US) for urban sound classification. (2) \emph{Speech} classification: Speech Commands v1 and v2 \cite{warden2018speech}  (SC1, SC2) for keyword spotting. (3)\emph{Music} classification: Openmic \cite{humphrey2018openmic} (OM) for multi-instrument recognition and GTZAN \cite{tzanetakis2002musical} for music genre classification. Note that we use the same training pipeline with \cite{gong21b_interspeech}, for ESC and SC2, we directly report the result of the paper \cite{gong21b_interspeech, gong2022ssast}. In particular, we follow the official train-test split. Since GZTAN does not have an official split, we followed the split provided by PyTorch source code\footnote{\url{https://pytorch.org/audio/stable/_modules/torchaudio/datasets/gtzan.html\#GTZAN}}. The specific experimental settings are as shown in Table \ref{tab: experimental setting}, We refer the readers to find more details in our source code repository\footnote{\url{https://github.com/MichaelLynn1996/AAT}}.

\begin{table}[h]
\renewcommand{\arraystretch}{1.2}
\caption{The specific settings for each dataset.}
    \label{tab: experimental setting}
    \resizebox{\linewidth}{!}{
    \begin{tabular}{c|cccccc}
\hline
\multirow{3}{*}{} & \multicolumn{6}{c}{Task}                                                                               \\ \cline{2-7} 
                  & \multicolumn{2}{c|}{\emph{Event}}         & \multicolumn{2}{c|}{\emph{Speech}}           & \multicolumn{2}{c}{\emph{Music}} \\ \cline{2-7} 
                  & ESC   & \multicolumn{1}{c|}{US}    & SC2     & \multicolumn{1}{c|}{SC1}    & GZTAN       & OM          \\ \hline
Class             & 50    & \multicolumn{1}{c|}{10}    & 30      & \multicolumn{1}{c|}{30}     & 10          & 20          \\
Scale             & 2,000 & \multicolumn{1}{c|}{8,732} & 105,829 & \multicolumn{1}{c|}{64,727} & 1,000        & 20,000      \\
Duration          & 5s    & \multicolumn{1}{c|}{4s}    & 1s      & \multicolumn{1}{c|}{1s}     & 30s         & 10s         \\
batch size        & 42    & \multicolumn{1}{c|}{48}    & 128     & \multicolumn{1}{c|}{128}    & 2           & 12          \\
Learning rate     & 1e-04 & \multicolumn{1}{c|}{1e-05} & 5e-04   & \multicolumn{1}{c|}{5e-04}  & 1e-5        & 1e-04       \\
epoch             & 25    & \multicolumn{1}{c|}{25}    & 30      & \multicolumn{1}{c|}{30}     & 30          & 30          \\ \hline
\end{tabular}
}
\end{table}

\begin{table*}[t]
\renewcommand{\arraystretch}{1.2}
\centering
\caption{Comparison with various fine-tuning methods, with the best results \textbf{highlighted}, excluding full fine-tuning for baseline. Besides, we also report the parameter percentage.}
  \label{tab: main result}
% \resizebox{0.9\linewidth}{!}{
\begin{tabular}{c|l|l|llllll}
\hline
\multicolumn{1}{l|}{\multirow{3}{*}{Model}}                            & \multicolumn{1}{c|}{\multirow{3}{*}{Method}} & \multicolumn{1}{c|}{\multirow{3}{*}{\begin{tabular}[c]{@{}c@{}}Tuning Param.\\ / Percentage\\ (M / \%)\end{tabular}}} & \multicolumn{6}{c}{Task}                                                                                                                                                                                                                                                                                                                                                                                                                                                 \\ \cline{4-9} 
\multicolumn{1}{l|}{}                                                  & \multicolumn{1}{c|}{}                        & \multicolumn{1}{c|}{}                                                                                              & \multicolumn{2}{c|}{\emph{Event}}                                                                                                                              & \multicolumn{2}{c|}{\emph{Speech}}                                                                                                                              & \multicolumn{2}{c}{\emph{Music}}                                                                                                                           \\ \cline{4-9} 
\multicolumn{1}{l|}{}                                                  & \multicolumn{1}{c|}{}                        & \multicolumn{1}{c|}{}                                                                                              & \multicolumn{1}{c}{\begin{tabular}[c]{@{}c@{}}ESC\\ Acc. (\%)\end{tabular}} & \multicolumn{1}{c|}{\begin{tabular}[c]{@{}c@{}}US\\ Acc. (\%)\end{tabular}} & \multicolumn{1}{c}{\begin{tabular}[c]{@{}c@{}}SC2\\ Acc. (\%)\end{tabular}} & \multicolumn{1}{c|}{\begin{tabular}[c]{@{}c@{}}SC1\\ Acc. (\%)\end{tabular}} & \multicolumn{1}{c}{\begin{tabular}[c]{@{}c@{}}GTZAN\\ Acc. (\%)\end{tabular}} & \multicolumn{1}{c}{\begin{tabular}[c]{@{}c@{}}OM\\ mAP\end{tabular}} \\ \hline
\multirow{6}{*}{\begin{tabular}[c]{@{}c@{}}AST\\ (SL)\end{tabular}}    & Full                                         & 87.295 / 100\%                                                                                                       & 95.6                                                                       & \multicolumn{1}{l|}{87.9}                                                  & 97.9                                                                       & \multicolumn{1}{l|}{97.7}                                                   & 84.8                                                                         & 95.8                                                                 \\ \cline{2-9} 
                                                                       & Head                                         & 0.04 / 0.02\%                                                                                                        & 94.1                                                                       & \multicolumn{1}{l|}{85.0}                                                    & 61.8                                                                       & \multicolumn{1}{l|}{63.2}                                                   & 77.9                                                                         & 56.8                                                                 \\
                                                                       & Partial                                      & 42.544 / 48.49\%                                                                                                     & 96.1                                                                       & \multicolumn{1}{l|}{88.5}                                                  & 96.8                                                                       & \multicolumn{1}{l|}{96.8}                                                   & 81.4                                                                         & 94.1                                                                 \\
                                                                       & Prompt                                       & 0.128 / 0.15\%                                                                                                       & 94.2                                                                       & \multicolumn{1}{l|}{80.9}                                                  & 91.8                                                                       & \multicolumn{1}{l|}{89.4}                                                   & 73.1                                                                         & 67.2                                                                 \\
                                                                       & $\mathrm{AAT^{M}}$                           & 3.567 / 3.91\%                                                                                                       & 96.1                                                                       & \multicolumn{1}{l|}{88.5}                                                  & 97.5                                                                       & \multicolumn{1}{l|}{97.1}                                                   & 82.4                                                                         & 98.1                                                                 \\
                                                                       & $\mathrm{AAT^{MS}}$                          & 7.118 / 7.51\%                                                                                                       & \textbf{96.4}                                                              & \multicolumn{1}{l|}{\textbf{88.7}}                                         & \textbf{97.6}                                                              & \multicolumn{1}{l|}{\textbf{97.2}}                                          & \textbf{83.1}                                                                & \textbf{98.7}                                                        \\ \hline
\multirow{6}{*}{\begin{tabular}[c]{@{}c@{}}SSAST\\ (SSL)\end{tabular}} & Full                                & 87.295 / 100\%                                                                                                       & 88.8                                                                       & \multicolumn{1}{l|}{83.1}                                                  & 98.0                                                                         & \multicolumn{1}{l|}{97.4}                                                   & 71.0                                                                           & 83.8                                                                 \\ \cline{2-9} 
                                                                       & Head                                         & 0.04 / 0.02\%                                                                                                        & 31.8                                                                       & \multicolumn{1}{l|}{40.4}                                                  & 25.7                                                                       & \multicolumn{1}{l|}{25.7}                                                   & 20.3                                                                         & 36.1                                                                 \\
                                                                       & Partial                                      & 42.544 / 48.49\%                                                                                                     & \textbf{85.0}                                                                & \multicolumn{1}{l|}{\textbf{80.8}}                                         & 97.3                                                                       & \multicolumn{1}{l|}{\textbf{98.8}}                                          & \textbf{64.8}                                                                & 70.6                                                                 \\
                                                                       & Prompt                                       & 0.128 / 0.15\%                                                                                                       & 46.9                                                                       & \multicolumn{1}{l|}{52.4}                                                  & 88.7                                                                       & \multicolumn{1}{l|}{92.3}                                                   & 29.7                                                                         & 55.0                                                                   \\
                                                                       & $\mathrm{AAT^{M}}$                           & 3.567 / 3.91\%                                                                                                       & 69.0                                                                         & \multicolumn{1}{l|}{69.0}                                                    & 96.2                                                                       & \multicolumn{1}{l|}{96.2}                                                   & 41.7                                                                         & 72.0                                                                   \\
                                                                       & $\mathrm{AAT^{MS}}$                          & 7.118 / 7.51\%                                                                                                      & 75.45                                                                      & \multicolumn{1}{l|}{77.8}                                                  & \textbf{97.4}                                                              & \multicolumn{1}{l|}{96.9}                                                   & 58.6                                                                         & \textbf{81.6}                                                        \\ \hline
\end{tabular}
% }
\end{table*}

% \vspace*{-10pt}
\subsection{Results and Analysis}

\subsubsection{The effectiveness of the designed modules}

% \noindent\textbf{The effectiveness of the designed modules.} 
We compare the performance of different fine-tuning approaches in Table \ref{tab: main result} with the backbones pre-trained via the SSL. Take fine-tuning on AST as an example, the results show that AAT comprehensively surpasses fine-tuning task-specific head, partial tuning, and Prompt tuning methods. Specifically, $\mathrm{AAT^{M}}$ outperforms partial tuning on acoustics benchmark ESC, SC2, SC1, GTZAN and OM, by 0.3\%, 0.8\%, 0.4\%, 1.2\% and 4.8\%, with the parameter count has been reduced by 83.3\%. Even in comparison with full fine-tuning, while achieving competitive performance, we can surpass full fine-tuning on benchmark ESC, US, and OM. For $\mathrm{AAT^{MS}}$, experimental results demonstrate that adding an additional Adapter for spatial adaptation can yield further performance improvements, validating the effectiveness of our theory.

\subsubsection{The impact of different pre-trained models}
% \noindent\textbf{The impact of different pre-trained Models.} 
From the results of fine-tuning on SSAST, we can first observe that, compared to fine-tuning the task-specific head and Prompt tuning methods yield conclusions similar to those obtained from fine-tuning on AST. However, even though our proposed AAT still achieves competitive performance, fine-tuning partial parameters on ESC, US, and SC1 yields better metrics, with full fine-tuning exhibiting the best performance. We believe the reason for this is that models trained with SSL methods extract features that are more general and task-agnostic, making it more challenging to transfer them to specific downstream tasks. It is worth noting that $\mathrm{AAT^{MS}}$ achieves a more significant performance improvement in fine-tuning on SSAST compared to $\mathrm{AAT^{M}}$. This further underscores the effectiveness of adding a Spatial Adapter.
% and the robustness of transferring general features with MLP Adapter.

% \vspace*{-100pt}
\begin{table*}[t]
\renewcommand{\arraystretch}{1.2}
\centering
\caption{Result of the composition of $\mathrm{AAT^{MS}}$ and Prompt tuning, comparing with $\mathrm{AAT^{MS}}$ and the best results \textbf{highlighted}.}
  \label{tab: joint result}
% \resizebox{0.9\linewidth}{!}{
\begin{tabular}{c|c|c|cccccc}
\hline
\multirow{3}{*}{Model}                                                 & \multirow{3}{*}{Method} & \multirow{3}{*}{\begin{tabular}[c]{@{}c@{}}{\begin{tabular}[c]{@{}c@{}}Tuning Param.\\ / Percentage\\ (M / \%)\end{tabular}}\end{tabular}} & \multicolumn{6}{c}{Task}                                                                                                                                                                                                                                                                                                                                                                 \\ \cline{4-9} 
                                                                       &                         &                                                                             & \multicolumn{2}{c|}{\emph{Event}}                                                                                                          & \multicolumn{2}{c|}{\emph{Speech}}                                                                                                          & \multicolumn{2}{c}{\emph{Music}}                                                                                   \\ \cline{4-9} 
                                                                       &                         &                                                                             & \begin{tabular}[c]{@{}c@{}}ESC\\ Acc. (\%)\end{tabular} & \multicolumn{1}{c|}{\begin{tabular}[c]{@{}c@{}}US\\ Acc. (\%)\end{tabular}} & \begin{tabular}[c]{@{}c@{}}SC2\\ Acc. (\%)\end{tabular} & \multicolumn{1}{c|}{\begin{tabular}[c]{@{}c@{}}SC1\\ Acc. (\%)\end{tabular}} & \begin{tabular}[c]{@{}c@{}}GTZAN\\ Acc. (\%)\end{tabular} & \begin{tabular}[c]{@{}c@{}}OM\\ mAP\end{tabular} \\ \hline
\multirow{2}{*}{\begin{tabular}[c]{@{}c@{}}AST\\ (SL)\end{tabular}}    & $\mathrm{AAT^{MS}}$     & 7.118 / 7.51\%                                                              & \textbf{96.4}                                          & \multicolumn{1}{c|}{\textbf{88.7}}                                         & \textbf{97.6}                                          & \multicolumn{1}{c|}{\textbf{97.2}}                                          & \textbf{83.1}                                            & \textbf{98.7}                                    \\
                                                                       & Joint                   & 7.236 / 7.69\%                                                              & 95.7                                                   & \multicolumn{1}{c|}{87.9}                                                  & 97.4                                                   & \multicolumn{1}{c|}{97.2}                                                   & 80.7                                                     & 97.8                                             \\ \hline
\multirow{2}{*}{\begin{tabular}[c]{@{}c@{}}SSAST\\ (SSL)\end{tabular}} & $\mathrm{AAT^{MS}}$     & 7.118 / 7.51\%                                                              & 75.45                                                  & \multicolumn{1}{c|}{\textbf{77.8}}                                         & 97.4                                                   & \multicolumn{1}{c|}{96.9}                                                   & \textbf{58.6}                                            & 81.6                                             \\
                                                                       & Joint                   & 7.236 / 7.69\%                                                              & \textbf{76.8}                                          & \multicolumn{1}{c|}{75.4}                                                  & \textbf{97.5}                                          & \multicolumn{1}{c|}{\textbf{97.1}}                                          & 55.2                                                     & \textbf{82.3}                                    \\ \hline
\end{tabular}
% }
\end{table*}

\subsection{Composition of AAT and Prompt tuning}

We also attempted to combine $\mathrm{AAT^{MS}}$ with Prompt tuning, forming \textbf{Joint} tuning. The experimental results for this are shown in Table \ref{tab: joint result}. As a result, we observed an interesting finding. During joint tuning on AST, its performance, while surpassing Prompt tuning on its own, did not reach the $\mathrm{AAT^{MS}}$ metrics. Meanwhile, on SSAST, four datasets showed improved performance with joint tuning compared to $\mathrm{AAT^{MS}}$, while the other two were below the level achieved by $\mathrm{AAT^{MS}}$. It appears that Prompt tuning has a somewhat inhibitory effect on the transfer of AAT to SL pre-trained AST. 

We analyze this because the features extracted from models pre-trained with SL contain a significant amount of task-specific information, and Prompt tuning operates at the feature dimension level, making it challenging to be effective. As analyzed earlier, features extracted from models pre-trained with SSL are more general and task-agnostic, which is why Prompt tuning can have a certain effect in those cases. We do not dismiss the possibility of combining various PEFT methods. Further exploration is warranted to determine the optimal combination of PEFT methods based on the model and the specific task, which could yield greater benefits.

% \vspace*{-4pt}
\subsection{Training Time}

The time consumed per epoch of different methods was counted, as shown in Table \ref{tab: training time}. We trained our model on two NVIDIA RTX 3090 graphic cards. Due to various factors that might affect training time, we report approximate results recorded in the log. The AAT is 8.9\% relatively faster than the full fine-tuning. We believe that the AAT may be a practical choice to fine-tune a pre-trained audio Transformer, which trains faster with less GPU memory (less trainable parameters) and performs competitively with full fine-tuning.

\begin{table}[h]
\renewcommand{\arraystretch}{1.2}
\caption{Consumed time per epoch of each method (on the speech command v1).}
    \label{tab: training time}
    \resizebox{\linewidth}{!}{
\begin{tabular}{ccccccc}
\hline
\multirow{2}{*}{Method}                                                  & \multirow{2}{*}{Full} & \multirow{2}{*}{Head} & \multirow{2}{*}{Partial} & \multirow{2}{*}{Prompt} & \multirow{2}{*}{$\mathrm{AAT^{M}}$} & \multirow{2}{*}{$\mathrm{AAT^{MS}}$} \\
                                                                         &                       &                       &                          &                         &                                     &                                      \\ \hline
\multirow{2}{*}{\begin{tabular}[c]{@{}c@{}}Time\\ (second)\end{tabular}} & \multirow{2}{*}{146}  & \multirow{2}{*}{60}   & \multirow{2}{*}{91}      & \multirow{2}{*}{92}      & \multirow{2}{*}{133}                & \multirow{2}{*}{134}                 \\
                                                                         &                       &                       &                          &                         &                                     &                                      \\ \hline
\end{tabular}
}
\end{table}

% We evaluate our method on the same five commonly used datasets as previous work, which provide a variety of downstream audio and speech challenges. We evaluate audio event classification performance by fine-tuning on the ESC-50 \cite{piczak2015dataset} (ESC) datasets. ESC-50 tests the acoustics classification of the environmental audio. Like the SSAST paper, we fine-tune on only the balanced partition of AudioSet and test on the eval partition. We evaluate speech performance by fine-tuning on Speech Commands 1 and 2 \cite{warden2018speech} for the task of keyword spotting (KS1, KS2), VoxCeleb \cite{NAGRANI2020101027} for speaker identification (SID), and IEMOCAP \cite{busso2008iemocap} for emotion recognition (ER). 

% To make a fair comparison with the previous work's full fine-tuning result, for ESC and KS2, we use the fine-tuning pipeline from AST. Specifically, we use mixup training \cite{DBLP:conf/iclr/TokozumeUH18}, SpecAugment \cite{Park2019}, an initial learning rate of 1e-4, and 2.5e-4, batch size of 42, and 128 and train the model with 25, and 30 epochs for ESC, KS2, respectively. For KS1, SID, and ER we use the SUPERB \cite{yang21c_interspeech} framework, the same fine-tune setting as SSAST. We use default settings for all models, which matched the results of a learning rate grid search.
% \vspace*{-5pt}
\section{Conclusion and future work}
\label{sec:conclusion}

In this paper, we have introduced a PEFT method based on Adapter tuning, known as AAT, for adapting audio Transformers to various downstream acoustic tasks. AAT includes both MLP Adapter and Spatial Adapter. The MLP Adapter runs in parallel with the MLP layer of the Transformer Block, facilitating the effective fusion of task-specific and generic features. The Spatial Adapter is placed after the MHSA layer to handle spatial information variations caused by different sample lengths in various downstream tasks. Extensive experiments on six datasets have demonstrated the effectiveness of AAT and we conducted preliminary exploration into the effectiveness of combining AAT with Prompt tuning. In future work, we plan to explore the integration of AAT with other PEFT methods to unlock its potential in the field of acoustics recognition.

% References should be produced using the bibtex program from suitable
% BiBTeX files (here: strings, refs, manuals). The IEEEbib.bst bibliography
% style file from IEEE produces unsorted bibliography list.
% -------------------------------------------------------------------------
\bibliographystyle{IEEEbib}
% \ninept
\bibliography{manuscript_final}

\end{document}